\documentclass[twocolumn,english]{IEEEtran}
\usepackage[T1]{fontenc}
\usepackage{babel}
\usepackage{graphicx}
\usepackage{epsf}
\usepackage[unicode=true,
 bookmarks=true,bookmarksnumbered=true,bookmarksopen=true,bookmarksopenlevel=1,
 breaklinks=false,pdfborder={0 0 0},backref=false,colorlinks=false]
 {hyperref}
\hypersetup{pdftitle={Your Title},
 pdfauthor={Your Name},
 pdfpagelayout=OneColumn, pdfnewwindow=true, pdfstartview=XYZ, plainpages=false}
\usepackage{breakurl}

\makeatletter
\ifCLASSOPTIONcompsoc
\usepackage[caption=false,font=normalsize,labelfont=sf,textfont=sf]{subfig}
\else
\usepackage[caption=false,font=footnotesize]{subfig}
\fi

\makeatother

\begin{document}

\title{Random load fluctuations and collapse probability of a power system
operating near codimension 1 saddle-node bifurcation}

\author{Dmitry Podolsky and Konstantin~Turitsyn,~\IEEEmembership{Member,~IEEE}%
\thanks{D. Podolsky and K. Turitsyn are with the Department of Mechanical
Engineering, Massachusetts Institute of Technology, Cambridge, MA,
02139. E-mails: \protect\href{http://podolsky@mit.edu}{podolsky@mit.edu},
\protect\href{http://turitsyn@mit.edu}{turitsyn@mit.edu}.%
}}
\maketitle
\begin{abstract}
For a power system operating in the vicinity of the power transfer
limit of its transmission system, effect of stochastic fluctuations
of power loads can become critical as a sufficiently strong such fluctuation
may activate voltage instability and lead to a large scale collapse
of the system. Considering the effect of these stochastic fluctuations
near a codimension 1 saddle-node bifurcation, we explicitly calculate
the autocorrelation function of the state vector and show how its
behavior explains the phenomenon of critical slowing-down often observed
for power systems on the threshold of blackout. We also estimate the
collapse probability/mean clearing time for the power system and construct
a new indicator function signaling the proximity to a large scale
collapse. The new indicator function is easy to estimate in real time
using PMU data feeds as well as SCADA information about fluctuations
of power load on the nodes of the power grid. We discuss control strategies
leading to the minimization of the collapse probability.\end{abstract}

\section{Introduction}

\IEEEPARstart{I}{t} is well known that small stochastic fluctuations
of power load, although usually negligible in the vicinity of a stable
operating point, may potentially lead to a large scale cascading failure
if the power system operates close to a saddle-node bifurcation point
\cite{Canizares2002,Andersson2005}. Early detection and mitigation
of such failures is a problem of utmost importance in contemporary
world of constantly increasing power demand, where power grids often
operate in a precritical regime. 

Statistical properties of aggregated power load, their influence on
static and dynamic characteristics of power systems remain the subject
of extensive studies for the last three decades, see \cite{Canizares2002}
for the review. A very considerable attention has been given to stability
analysis of power networks based on Lyapunov theory of dynamical systems,
as it was recognized early that the proximity of an operating point
to saddle-node and/or Hopf bifurcations of power systems signals about
approach to an instability of the base state \cite{Kwatny1995}. In
particular, saddle-node bifurcations were associated to the phenomenon
of voltage collapse as well as loss of synchronism \cite{Taylor1994}.
Correspondingly, a multitude of stability criteria based on the estimation
of global Lyapunov functions of power systems have been introduced
(see for example classical papers \cite{DeMarco1987,DeMarco1990,Nwankpa1993}),
many corresponding indicator functions being already used in industry
for preventive control and dynamic security assessment of power grids
\cite{Canizares2002}. Despite its popularity, energy function analysis
of power systems is not entirely free of drawbacks: (a) typically,
a knowledge of the values of system variables on all nodes of the
network is needed in order to estimate the global Lyapunov function
in a given operating regime, and (b) even if such tremendous amount
of data is available in real time, the estimation process has a high
computational cost, which becomes especially critical in the proximity
of a large scale collapse.

Assuming in the present work that the operating point of the power
system under consideration is close to a saddle-node bifurcation,
we extend the standard approach to Lyapunov stability analysis of
power systems, taking into account that for a typical power grid without
any specific structural symmetries the center manifold is one-dimensional,
i.e., saddle-node bifurcation has codimension 1. This observation
allows us to explicitly calculate the autocorrelation function of
system variables in the operating regime near the bifurcation point
and find an approximate expression for the mean clearing time/probability
of a large scale failure of the power system. Using the latter, we
construct a new indicator function of proximity to the voltage collapse/loss
of synchronism, which is significantly easier to estimate in real
time than the global Lyapunov function, especially if the system operator
receives real time PMU data as well as SCADA information about fluctuations
of power demand on individual nodes.

This manuscript is organized as follows. In Sec. \ref{sec:Structure-preserving-model}
we review the structure-preserving model used in the present paper
to describe dynamic behavior of system variables in the vicinity of
the saddle-node bifurcation point. The autocorrelation function of
system variables is explicitly calculated in the Sec. \ref{sec:General-case-autocorrelation},
while a new estimate for the mean clearing time is given in the Sec.
\ref{sec:General-case-mct}. In the Sec. \ref{sec:Numerical-simulations}
we perform the validation of our model by checking the derived formulae
against numerical simulations of the IEEE 39 bus (New England) power
system described in \cite{Pai}. Finally, the Sec. \ref{sec:Conclusion}
is devoted to discussion of control strategies for minimizing the
collapse probability and conclusions.

\section{The model\label{sec:Structure-preserving-model}}

To describe dynamics of system variables in the vicinity of a saddle
node bifurcation point, we use the system of coupled swing equations
on $(P,V)$ nodes of the grid 
\begin{equation}
\frac{H_{i}}{\pi f_{0}}\frac{d^{2}\theta_{i}}{dt^{2}}+\alpha_{i}\frac{d\theta_{i}}{dt}=\sum_{j\sim i}\mathcal{Y}_{ij}V_{i}V_{j}\sin(\theta_{i}-\theta_{j}-\gamma_{ij})+P_{m,i}\label{eq:swing}
\end{equation}
and power flow equations on $(P,Q)$ nodes
\begin{equation}
P_{0,i}+\alpha_{p,i}\dot{\theta}_{i}+\beta_{p,i}V_{i}+T_{p,i}\dot{V}_{i}=\sum_{j\sim i}{\cal Y}_{ij}V_{i}V_{j}\sin(\theta_{i}-\theta_{j}-\gamma_{ij}),\label{eq:P}
\end{equation}
\begin{equation}
Q_{0,i}+\alpha_{q,i}\dot{\theta}_{i}+\beta_{q,i}V_{i}+T_{q,i}\dot{V}_{i}=\sum_{j\sim i}{\cal Y}_{ij}V_{i}V_{j}\cos(\theta_{i}-\theta_{j}-\gamma_{ij}),\label{eq:Q}
\end{equation}
where as usual $\theta_{i}$ is a voltage phase on a bus $i$, $H_{i}$
is an inertia constant for a generator on the node $i$, $\alpha_{i}$
denote frequency controls on the $(P,V)$ nodes with generators, $V_{i}$
is a voltage magnitude on a bus $i$ (for the $(P,V)$ nodes, $V_{i}=E_{i}$).
Real and reactive power loads on $(P,Q)$ nodes are generally (weakly
changing) functions of frequency $\dot{\theta}_{i}$, voltage $V_{i}$
and voltage change rate $\dot{V}_{i}$. If only dynamics of system
variables at the vicinity of a stable operating point is to be considered,
it is sufficient to keep first leading orders in the Taylor expansions
of real and reactive power loads in powers of their arguments. This
corresponds to the structure preserving model of \cite{Bergen1981}
extended to the case of load dependence on the voltage change rate
$\dot{V}_{i}$. 

The loads have a fixed power factor $k_{i}$. They fluctuate with
time, and fluctuations of loads on different nodes of the grid are
statistically independent. We assume that their correlation properties
are Gaussian:
\begin{equation}
\langle\delta P_{i}(t)\delta P_{j}(t')\rangle=B_{i}(t-t')\delta_{ij}.\label{eq:Pcorr}
\end{equation}
One simple example of the function $B_{i}$ is exponential $B_{i}(t-t')=A_{i}\exp(-|t-t'|/\tau_{i})$,
where the characteristic time scale $\tau$ is the decay time of temporal
correlations of power load fluctuations.

Since the load power factor is fixed, fluctuations of reactive loads
are related to (\ref{eq:Pcorr}) as 
\[
\langle\delta Q_{i}(t)\delta Q_{j}(t')\rangle=\frac{k_{i}^{2}-1}{k_{i}^{2}}B_{i}(t-t')\delta_{ij},
\]
so that only fluctuations of $\delta P_{i}$ are needed to be considered. 

Note that $(\ref{eq:Pcorr})$ describes the autocorrelation function
of \emph{a stationary process}. The property of stationarity (or translational
invariance in time) $t\to t+\delta t$, $t'\to t'+\delta t$ is only
approximate, as behavior of loads and their random fluctuations features
natural cycles (for example, day/night). However, we are only interested
to study a short time scale (tens of seconds and minutes) behavior
of the state vector, when (\ref{eq:Pcorr}) is perfectly applicable. 

After solving the power flow equations, finding the base state and
linearizing equations (\ref{eq:swing}), (\ref{eq:P}), (\ref{eq:Q})
about the stable operating point, one finds the matrix stochastic
differential equations (SDE)
\begin{equation}
{\cal M}\ddot{x}+{\cal D}\dot{x}+{\cal K}x={\cal A}x=\delta P.\label{eq:swinglinearized}
\end{equation}
Here $x$ is the system state vector including voltage phases and
magnitudes, the matrix ${\cal M}$ describes inertial properties of
generators connected to the grid, the matrix ${\cal D}$ --- primary
frequency controls on $(P,V)$ nodes as well as frequency and $\dot{V}$
dependence of power loads on $(P,Q)$ nodes, and finally ${\cal K}$
is the power flow Jacobian encoding all static properties of the power
system. The system (\ref{eq:swinglinearized}) of SDE will be the
main subject of our study.

\section{Calculation of autocorrelation function\label{sec:General-case-autocorrelation}}

Generally, the autocorrelation function of the system vector $x$
can be found as
\begin{equation}
\langle x(t)x^{T}(t')\rangle={\rm Re}\int\frac{d\omega}{2\pi}e^{-j\omega(t-t')}\langle x(\omega)x^{\dagger}(\omega)\rangle,\label{eq:CorrFourier}
\end{equation}
where 
\[
\langle x(\omega)x^{T}(-\omega)\rangle=\mathcal{A}^{-1}(\omega)\mathcal{B}(\omega)({\cal {\cal A}}^{\dagger}(\omega))^{-1}
\]
and the Fourier-transformed system matrix is
\[
\mathcal{A}(\omega)=-\mathcal{M}\omega^{2}+j\mathcal{D}\omega+\mathcal{K}.
\]
The correlation properties of fluctuating loads are given by the matrix
\[
\mathcal{B}(\omega)=\langle\delta P(\omega)\delta P^{\dagger}(\omega)\rangle=B(\omega)\mathbf{1}.
\]
The value of the integral (\ref{eq:CorrFourier}) is determined by
the singularities of the integrand in the complex $\omega$ plane,
which in particular include zeros of $\det{\cal A}(\omega)$ and $\det{\cal A}(-\omega)$,
as well as the singularities of ${\cal B}(\omega)$. In the regime
of interest, dynamics of system variables is largely dictated by very
slow modes (see below). Power loads fluctuate rapidly compared to
the time scales of this slow dynamics, their correlation in time described
by (\ref{eq:Pcorr}) becomes negligible, and one can simply write
${\cal B}(\omega)={\cal B}\cdot\mathbf{1}$, where $B_{ij}=\int_{0}^{+\infty}dtB_{ij}(t)$.
For the case $B_{ij}(t)=A_{i}\exp(-t/\tau_{i})\delta_{ij}$ considered
in the previous Section, one expect that the time scales $\tau_{i}$
are the shortest in the system, and one effectively has $B_{ij}=A_{i}/\tau_{i}\delta_{ij}$.

Naturally, the singularity closest to the real $\omega$ axis determines
behavior of the autocorrelation function (\ref{eq:CorrFourier}) at
late times. To identify it, we note that at a saddle-node bifurcation
point several eigenvalues of the power flow Jacobian ${\cal K}$ vanish.
There exists only one such vanishing eigenvalue $\epsilon\to0$ if
the center manifold of the power system is one-dimensional \cite{Dobson1992a}.
Performing the eigenvalue decomposition of the inverse power flow
Jacobian
\[
{\cal K}^{-1}=\sum_{i}b_{i}\Lambda_{i}^{-1}a_{i}^{T},
\]
where $\Lambda_{i}$ are eigenvalues of ${\cal K}$ and $a_{i}$ and
$b_{i}$ are corresponding left and right eigenvectors, we see that
if the operating point of the power system is close to the saddle-node
bifurcation point, ${\cal K}^{-1}$ is approximately given by 
\begin{equation}
{\cal K}^{-1}=\frac{1}{\epsilon}ba^{T}+\ldots,\label{eq:Krepresent}
\end{equation}
where $\epsilon$ is the eigenvalue of ${\cal K}$ vanishing at the
bifurcation point. Dots denote the contribution of all the other eigenvalues
of ${\cal K}$, which is at most of the order ${\cal O}(1)$ in powers
of $\epsilon$ (although for large systems it can be numerically large).
We are particularly interested in the situation when the 2-norm $||\frac{1}{\epsilon}ba-{\cal K}^{-1}||_{2}\ll||{\cal K}^{-1}||_{2}$
and the asymptotic representation (\ref{eq:Krepresent}) holds well. 

In the case under consideration, the leading singularity of the integrand
in (\ref{eq:CorrFourier}) coincides with a zero of $\det{\cal A}(\omega)$
(or $\det{\cal A}(-\omega)$ depending on the sign of the time difference
$t-t'$). Such singularity is a simple pole by assumption that the
center manifold of the power system is one-dimensional. Solving the
equation $\det{\cal A}(\omega)=\det(-{\cal M}\omega^{2}+i{\cal D}\omega+{\cal K})=0$
by means of perturbation theory in the small parameter $\epsilon$,
one finds that the mode determining behavior of the autocorrelation
function at late times is given by
\begin{equation}
\omega_{0}\approx-\frac{j}{{\rm Tr}({\cal D}{\cal K}^{-1})}=-\frac{j\epsilon}{a^{T}{\cal D}b}+{\cal O}(\epsilon^{2}).\label{eq:s0}
\end{equation}
The frequency of the leading mode is purely imaginary.%
\footnote{An oscillating contribution disappears from the mode for sufficiently
small $\epsilon<(a^{T}{\cal D}b)^{2}/4a^{T}{\cal M}b$.%
} 

Estimating the integral (\ref{eq:CorrFourier}) in the vicinity of
this leading singularity, one finally finds (to the leading order
in $\epsilon$)
\begin{equation}
\langle x(0)x^{T}(t)\rangle=\frac{b(a^{T}{\cal B}a)b^{T}}{2\epsilon a^{T}{\cal D}b}\exp\left(-\frac{\epsilon t}{a^{T}{\cal D}b}\right).\label{eq:CorrFfinal}
\end{equation}
Note that in this operating regime near the codimension 1 saddle-node
bifurcation point \emph{dynamics} of system variables $x(t)$ is directly
related to \emph{static }characteristics of the power system such
as the lowest eigenvalue $\epsilon$ of the power flow Jacobian. 

As follows from (\ref{eq:CorrFfinal}), the right eigenvector $b$
of the power flow Jacobian ${\cal K}$ determines the preferred direction
in the phase space of a power system near a codimension 1 saddle-node
bifurcation \cite{Dobson1992a,Canizares1995}. In turn, the left eigenvector
$a$ shows fluctuations of the loads on which nodes mostly determine
stochastic dynamics of the state vector: the nodes $i$ with larger
components $a_{i}$ are the ones, where fluctuations of the power
loads $\delta P_{i}$, $\delta Q_{i}$ influence the dynamics of the
state vector stronger. 

As the operating point of the power system approaches the power transfer
limit and the lowest eigenvalue $\epsilon$ of the power flow Jacobian
gets smaller, the rate of decay of the autocorrelation function (\ref{eq:CorrFfinal})
decreases quickly. This observation is known as the phenomenon of
critical slowing-down observed during large scale failures in power
grids \cite{Hines2011,Cotilla-Sanchez2012}. 

Finally, we would also like to emphasize that fluctuations of the
leading mode, the component of the system vector $x$ along $b$ direction,
become strongly amplified as $\epsilon$ decreases, since the amplitude
of the leading mode near the bifurcation is proportional to $\epsilon^{-1/2}$.
One of the main goals of system control in this regime is to decrease
the amplitude of the leading mode by adjusting the values of the matrix
element $a^{T}{\cal D}b$ (and, if possible, $a^{T}{\cal B}a$). This
can be done using primary frequency control on $(P,V)$ nodes as well
as utilizing resources of load following on $(P,Q)$ nodes as we discuss
below.

\section{Mean clearing time and probability of voltage collapse\label{sec:General-case-mct}}

Taking the results of the previous Section into account, it is also
possible to explicitly calculate the mean clearing time for an operating
point close%
\footnote{But not too close as will become clear from the subsequent discussion.%
} to the bifurcation point. Namely, considering a scalar reduced system
variable $z(t)$ defined according to
\begin{equation}
z(t)=b^{T}x(t),\label{eq:cdef}
\end{equation}
one finds the following equation of motion for it:
\begin{equation}
a^{T}{\cal M}b\cdot\ddot{z}+a^{T}{\cal D}b\cdot\dot{z}+\epsilon z+a^{T}\Gamma bb\cdot z^{2}+\ldots=a^{T}\delta P,\label{eq:SingleSystemNL}
\end{equation}
where the matrix $\Gamma_{ijk}=\frac{\partial{\cal K}_{ij}}{\partial\theta_{k}}$
is the first derivative of the power flow Jacobian w.r.t. the system
variables and $\ldots$ denote higher derivatives of ${\cal K}$.
The equation (\ref{eq:SingleSystemNL}) is a scalar SDE describing
the process of activation of an over-damped unharmonic oscillator.
Therefore, one can simply apply the classical result by Kramers \cite{Kramers1940}
to calculate the probability of collapse/the mean clearing time: 
\begin{equation}
t_{{\rm mct}}\approx\frac{2\pi a^{T}{\cal D}b}{\epsilon}\exp\left(\frac{\epsilon^{3}a^{T}{\cal D}b}{3(a^{T}\Gamma bb)^{2}(a^{T}{\cal B}a)}\right)\label{eq:mct}
\end{equation}
Estimates similar to this one were previously found using Lyuapunov
stability analysis of power systems (see for example \cite{DeMarco1990,Nwankpa1993}).
Yet, the expression (\ref{eq:mct}) is different from these well-known
results in one important respect: the mean clearing time (\ref{eq:mct})
depends only on a small number of parameters of the power system,
unlike the global Lyuapunov function which is a functional of dynamical
characteristics of the power system. This is so because the number
of degrees of freedom of a power system gets enormously reduced in
the vicinity of bifurcation, and all the information about the power
system in the vicinity of the bifurcation point is aggregated through
a small number of quantities. These quantities (such as matrix elements
$a^{T}{\cal D}b$ and $a^{T}{\cal B}a$ or the lowest eigenvalue $\epsilon$
of the power flow Jacobian) entering (\ref{eq:mct}) can be straightforwardly
estimated from the static state analysis using the measurements of
the correlation function $\langle x(0)x^{T}(t)\rangle$ of the state
vector (containing the information about the vector $b$, the matrix
element $a^{T}{\cal D}b$ and $\epsilon$) and local SCADA data of
power consumption on individual nodes of the grid (which encode information
about the matrix ${\cal B}$) and therefore do not require \emph{any
a priori knowledge} of the power system structure. Such measurements
will make it easier to estimate the mean clearing time for the power
system in real time using the expression (\ref{eq:mct}).

We also emphasize that the result (\ref{eq:mct}) is strictly speaking
applicable only to the case of power systems with one-dimensional
center manifold, and an estimate similar to (\ref{eq:mct}) would
be impossible to make if the dimension of center manifold of the power
system is 2 or larger. What is necessary for the derivation of (\ref{eq:mct})
is the equilibration law known in physics as the principle of detailed
balance, and it does not apply for system matrices of generic power
grids. However, because of tremendous reduction of the total number
of degrees of freedom in the vicinity of a codimension 1 bifurcation
point, the power system becomes effectively one-dimensional (there
is only one relevant degree of freedom), and the principle of detailed
balance is always valid for one-dimensional systems. 

\section{Numerical simulation of IEEE 39 power system\label{sec:Numerical-simulations}}

We shall now perform the verification of the model described in the
Section \ref{sec:Structure-preserving-model} comparing the expression
(\ref{eq:CorrFfinal}) for the correlation function with results of
numerical simulations of the IEEE 39 (New England) power system. The
strategy is as follows. First, the saddle-node bifurcation point is
localized using continuation power flow procedure \cite{Ajjarapu1992}
implemented in PSAT Toolbox for Matlab \cite{PSAT}, then it is identified
more precisely using MATPOWER 4.1 library for Matlab \cite{MATPOWER}.
An operating point close to the bifurcation is chosen, static power
flow equations --- solved using MATPOWER, the power flow Jacobian
--- calculated, and finally stochastic differential equations of the
Section \ref{sec:Structure-preserving-model} are solved using the
implicit Euler algorithm. The number $N_{r}$ of realizations of the
random load fluctuations is 1000 (we consider only a stationary regime
where averaging the system variables $x(t)$ over realizations of
the statistical ensemble (\ref{eq:Pcorr}) of random loads is expected
to be equivalent to averaging of $x(t)$ over time).

The load parameter $\lambda=2.12$ is chosen (if $\lambda=1$ corresponds
to the operating point described in \cite{Pai}). This corresponds
to a $0.9$\% displacement in terms of the generated real power as
compared to the critical regime.%
\footnote{The total generated real power at $\lambda_{{\rm crit}}\approx2.13569843$
is approximately $13.67$ GW.%
} As usual, the bus 31 is a slack bus. The state vector includes voltage
phases on the nodes 1-30, 32-39 and voltage magnitudes --- on the
nodes 1-29. The same ${\cal M}$ matrix chosen as presented in \cite{Pai},
with the base frequency $f_{0}=60$ Hz. The ${\cal D}$ matrix is
diagonal with components $d=1$ p.u. corresponding to voltage phases
on $(P,Q)$ buses, $d=0.1$ p.u. corresponding to voltage magnitudes
on $(P,Q)$ buses and $d=10$ p.u. --- to phases on $(P,V)$ buses.
The loads are allowed to fluctuate only on the buses 3, 10 and 21,
with the same characteristic amplitudes of fluctuations $\sqrt{B}=0.1$
p.u. For our choice of $\lambda$ the smallest eigenvalue of the power
flow Jacobian ${\cal K}$ is $\epsilon\approx0.1811$ p.u, with the
ratio of 2-norms $||{\cal K}^{-1}-\epsilon^{-1}ba||_{2}/||{\cal K}^{-1}||_{2}\approx0.0717$.
Therefore, the expansion in powers of $\epsilon$ is feasible, albeit
$\epsilon$ is not too small by itself (only smallness of dimensionless
quantities dictates where the perturbation theory is applicable). 

\begin{figure}[htbp]
\includegraphics[scale=0.55]{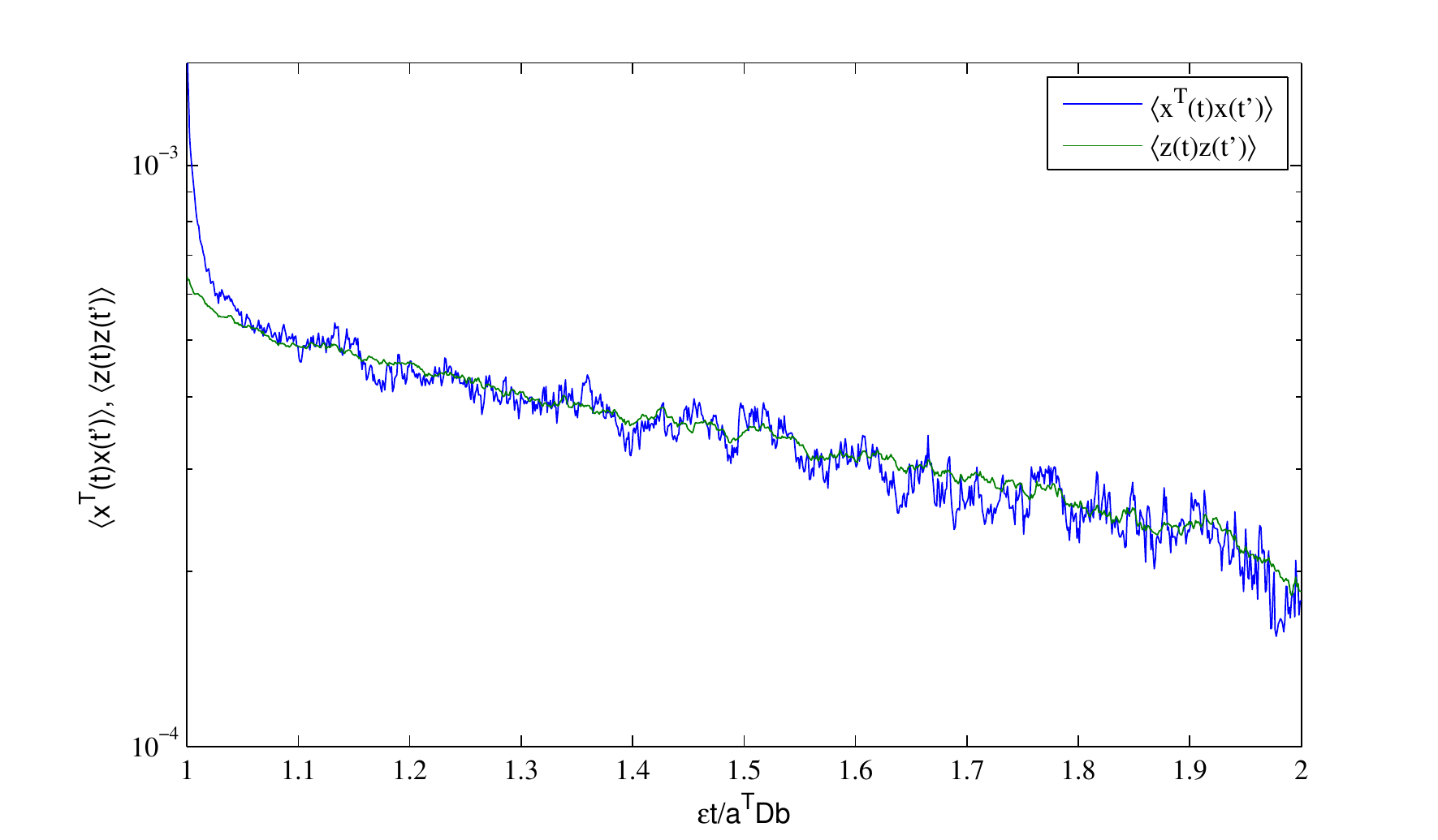}
\caption{Relation between the logarithms of the autocorrelation function $\langle z(t)z(t')\rangle$
of the reduced system variable $b^{T}x(t)$ (red) and the full autocorrelation
function $\langle x^{T}(t)x(t')\rangle$ (blue). The initial moment
of time $t$ is chosen to be $t=a^{T}{\cal D}b/\epsilon$. \label{fig:state-rec}}
\end{figure}

The numerical results for the autocorrelation function $\langle z(t)z(t')\rangle$
of the reduced system variable $z(t)=b^{T}x(t)$ as well as the scalar
autocorrelation function $\langle x^{T}(t)x(t')\rangle$ are presented
on the Fig. \ref{fig:state-rec}. The full correlation function of
the system vector $\langle x^{T}(t)x(t')\rangle$ clearly follows
the one of reduced variable $\langle z(t)z(t')\rangle$ except the
initial short interval of time, where contributions from sub-leading
modes into the autocorrelation function are not small. This confirms
our expectation that fluctuations of system variables orthogonal to
the direction of the vector $b$ become suppressed in the vicinity
of the saddle-node bifurcation. Note that for any given finite number
of realizations $N_{r}$ the autocorrelation function $\langle x^{T}(t)x(t')\rangle$
fluctuates stronger than $\langle z(t)z(t')\rangle$ because it also
accounts for fluctuations of the system vector orthogonal to the vector
$b$, although in the limit of infinite number of realizations both
correlation functions coincide with each other. 

For our choice of parameters the value of the amplitude of the correlation
function $\langle x^{T}(t)x(t')\rangle$ is $a^{T}{\cal B}a/2\epsilon a^{T}{\cal D}b\approx5.96\cdot10^{-4}$,
while the best fit value recovered from \ref{fig:state-rec} is $5.26\cdot10^{-4}$.
The autocorrelation decay time $\tau=a^{T}{\cal D}b/\epsilon\approx17.3$
sec also coincides well with the best fit value $\tau\approx17.0$
sec and an integral estimate for the average correlation time $\int_{0}^{t_{f}}dt\langle x^{T}(t)x(t')\rangle/\langle x^{T}(t)x(t)\rangle\approx18.1$
sec. 

\section{Discussion and path forward\label{sec:Conclusion}}

In the present paper we discuss a general power grid without any particular
structural symmetries operating near the power transfer limit of its
transmission system. A saddle-node bifurcation present in the phase
space of such power system typically has codimension 1, which implies
that dynamics of state variables near such bifurcation point is essentially
determined by a single degree of freedom which we denote as reduced
system variable. We consider how stochastic fluctuations of power
loads exactly influence this dynamics, explicitly calculating (a)
the autocorrelation function of the state vector (\ref{eq:CorrFfinal})
and (b) the probability of a large scale failure of the power system
in the vicinity of the saddle-node bifurcation point/the mean clearing
time for the system (\ref{eq:mct}). Although such an operating regime
was extensively studied in literature for the last 30 years (see \cite{Canizares2002}
and references therein), both results (a) and (b) are new to our knowledge.
Also, using (a) we are able to quantify the phenomenon of critical
slowing-down recently discovered in power grids operating on the threshold
of a large scale cascading failure \cite{Hines2011,Cotilla-Sanchez2012}.

Using the expression for the mean clearing time (\ref{eq:mct}), one
can introduce a new simple indicator function (\ref{eq:IndicatorFunction})
signaling proximity of a power system to a large scale failure. Indeed,
as follows from the expression (\ref{eq:mct}), collapse probability
is exponentially small when the expression 
\begin{equation}
I_{c}=\frac{\epsilon^{3}a^{T}{\cal D}b}{(a^{T}\Gamma bb)^{2}a^{T}{\cal B}a}\gg1\label{eq:IndicatorFunction}
\end{equation}
and vise versa. Even if the operating point of the power system under
consideration is close to the power transfer limit (i.e., $\epsilon$
nearly vanishes), the mean clearing time could still be exponentially
large when the matrix element $a^{T}{\cal D}b$ is large and/or the
matrix element $a^{T}{\cal B}a$ --- small. This in turn implies that
the probability of a large scale failure is low. 

As follows from (\ref{eq:IndicatorFunction}), a strategy for a system
operator to keep the power system close to the power transfer limit
of its transmission system while minimizing the probability of a large
scale collapse would be to maximally utilize resources of load following
\cite{Hirst1996} (effectively reducing the matrix element $a^{T}{\cal B}a$)
and/or adjust, if it is possible, parameters of the primary frequency
control on $(P,V)$ nodes (increasing the value of the matrix element
$a^{T}{\cal D}b$). Note that $a^{T}{\cal D}b$ cannot be increased
indefinitely: such increase for example would imply a growth of the
correlation time $\tau=a^{T}{\cal D}b/\epsilon$ leading in turn to
possible issues related to interference between primary and secondary
frequency controls.

There exist several reasons why the operating regime of a power system
discussed in the paper seems rather attractive despite the fact that
the operating point is located near the threshold of system instability.
First of all, in such a regime, the power grid and its transmission
system are utilized with the best effectiveness possible: the throughput
of the transmission system is maximal, nearly all energy which the
grid is capable to produce is consumed (assuming the minimal power
losses in the transmission system). Therefore, such operating regime
of the grid is optimal from the economical point of view given the
system remains under control and the operating costs of control systems
are not too high. 

Second, synchronism of the power system actually holds rather well
in the regime under consideration. Indeed, the degree of synchronism
is determined by the correlation function of the operating frequency
$\langle\delta\omega_{i}(t)\delta\omega_{j}(t')\rangle$, which for
the nodes $i$ with large representation in the vector $b$ is given
to the leading order in $\epsilon$ by 
\begin{equation}
\langle\delta\omega_{i}(t)\delta\omega_{j}(t')\rangle\approx\frac{\epsilon b_{i}(a^{T}{\cal B}a)b_{j}}{(a^{T}{\cal D}b)^{3}}\exp\left(-\frac{\epsilon|t-t'|}{a^{T}{\cal D}b}\right).\label{eq:CorrFreq}
\end{equation}
Control countermeasures necessary to maximize the value of the indicator
function (\ref{eq:IndicatorFunction}) and minimize the collapse probability
are the ones which also maximize the value of the matrix element $a^{T}{\cal D}b$
(and/or minimize $a^{T}{\cal B}a$) and therefore decrease the amplitude
of the correlation function (\ref{eq:CorrFreq}). 

Finally, since in the regime under consideration dynamics of the power
system is essentially described by a single degree of freedom, the
complicated problem of state recovery can be effectively solved by
using real time data feeds from synchrophasor measurement units as
well as a much larger (and much slower acquired) volume of SCADA data
about fluctuating aggregated power loads. The same applies to the
problem of system identification \cite{Ljung1998}: as we have discussed,
measurements of the autocorrelation function $\langle x(0)x^{T}(t)\rangle$
using PMU data feeds directly provide information about the matrix
elements $a^{T}{\cal D}b$ and $a^{T}{\cal B}a$, the lowest eigenvalue
of the power flow Jacobian, corresponding left and right eigenvectors
$a$ and $b$. Therefore, such measurements can potentially allow
for the approximate recovery of the system matrix in the vicinity
of the bifurcation point or at least its part responsible for the
dynamics of the leading mode of the state vector. We shall discuss
this observation more extensively elsewhere.

\section*{Acknowledgments}

This work was partially supported by NSF award ECCS-1128437 and MIT/SkTech
seed funding grant.


\end{document}